\documentclass[preprint,superscriptaddress,aps,pre]{revtex4}
\usepackage{graphicx,amsmath} 

\usepackage{graphicx}

\date{\today}

\begin{document}

\title{
Multiscale enhanced path sampling based on 
the Onsager-Machlup action: Application to a model polymer
}

\author{Hiroshi Fujisaki}\email{fujisaki@nms.ac.jp} 
\affiliation{
Department of Physics,
Nippon Medical School,
2-297-2 Kosugi-cho, Nakahara, Kawasaki 
211-0063, Japan
}
\affiliation{
Molecular Scale Team, 
Integrated Simulation of Living Matter Group,
Computational Science Research Program,
RIKEN, 2-1 Hirosawa, Wako 351-0198, Japan
}
\author{Motoyuki Shiga}\email{shiga.motoyuki@jaea.go.jp} 
\affiliation{
Center for Computational Science and E-Systems
Japan Atomic Energy Agency
5-1-5, Kashiwanoha, Kashiwa, Chiba, 277-8587, Japan
}
\author{Kei Moritsugu}\email{moritsuguk@riken.jp} 
\affiliation{
Molecular Scale Team, 
Integrated Simulation of Living Matter Group,
Computational Science Research Program,
RIKEN, 2-1 Hirosawa, Wako 351-0198, Japan
}
\affiliation{
Department of Medical Life Science, Graduate School of Medical Life Science,
Yokohama City University
1-7-29 Suehiro-cho, Tsurumi, 
Yokohama 230-0045, Japan
}
\author{Akinori Kidera}\email{kidera@tsurumi.yokohama-cu.ac.jp} 
\affiliation{
Molecular Scale Team, 
Integrated Simulation of Living Matter Group,
Computational Science Research Program,
RIKEN, 2-1 Hirosawa, Wako 351-0198, Japan
}
\affiliation{
Department of Medical Life Science, 
Graduate School of Medical Life Science,
Yokohama City University
1-7-29 Suehiro-cho, Tsurumi, 
Yokohama 230-0045, Japan
}

\begin{abstract}


We propose a novel path sampling method 
based on the Onsager-Machlup (OM) action by
generalizing the multiscale enhanced sampling (MSES) technique suggested
by Moritsugu and coworkers (J.~Chem.~Phys.~{\bf 133}, 224105 (2010)).
The basic idea of this method is 
that the system we want to study (for example, some molecular system 
described by molecular mechanics) 
is coupled to a coarse-grained (CG) system,
which can move more quickly and computed more efficiently than 
the original system. We simulate this combined system 
(original + CG system) using (underdamped) Langevin dynamics 
where different heat baths are coupled to the two systems. 
When the coupling is strong enough, the 
original system is guided by the CG system, and able to 
sample the configuration and path space more efficiency.
We need to correct the bias caused by the coupling, however,
by employing the Hamiltonian replica exchange where we 
prepare many path replica with different coupling strengths.
As a result, an unbiased path ensemble for the original system
can be found in the weakest coupling path ensemble.
This strategy is easily implemented because a weight for a path 
calculated by the OM action is formally the same as the Boltzmann weight
if we properly define the path ``Hamiltonian''. 
%
%
%
We apply this method to a model polymer with Asakura-Oosawa interaction,
 and compare the results with the conventional transition path sampling
method.

\end{abstract}

\maketitle

\section{Introduction}




Molecular dynamics (MD) simulation \cite{MDbook1,MDbook2} is a rigorous 
and versatile approach
for investigating dynamic trajectory of molecular systems by numerical
integration of Newton's equation.
Its applicability has been quickly expanded 
in time and length scales
owing to the development of
hardwares (general purpose ones such as PC clusters, GPGPU, supercomputers,
or special purpose ones such as Anton \cite{Anton}), convenient softwares
with reliable force fields (Amber \cite{amber}, Gromacs \cite{gromacs}, 
CHARMM \cite{charmm}, NAMD \cite{namd} and others),
 and efficient numerical algorithms \cite{MDbook2}.
However, because the MD step size should be close to the 
vibrational time scales
of the molecular system ($\sim$ 1fs),
conventional MD simulations have not been applicable to 
processes that are extremely rare,
slow, and/or diffusive processes.
%
While the MD simulation is able to explore fast structural fluctuations
around a basin of potential (or free) energy surface, it often 
fails to sample interbasin hopping, which describes
various phenomena such as protein folding and chemical 
reactions \cite{TPS,TPS2}.


To overcome this problem, several methods have been
proposed and tested over the last two decades 
(for reviews, 
see \cite{TPS3,FFMK09}). 
One of the pioneering works 
is transition path sampling (TPS) 
developed by Chandler, Dellago, Bolhuis and coworkers \cite{TPS,TPS2,TPS3}.   
Since then, there has been effort to improve the efficiency of TPS calculations,
such as transition interface sampling (TIS) \cite{TIS} and partial 
TIS \cite{PPTIS}.
%
Related to these methods are 
milestoning \cite{milestone} and forward flux sampling \cite {FFsampling}.
Studies on further improvement of the efficient path sampling is still
under way.



An alternative approach to path sampling is the action-based methods 
such as the Onsager-Machlup (OM) action method \cite{Wiegelbook,Zuckermanbook}.
The OM action has been used for path search problems by Elber 
and coworkers \cite{EGC02}, 
Eastman and coworkers \cite{EGD01}, 
and Orland and coworkers \cite{Orland09}.
In the early works, this method has been used to find the 
most dominant pathway, 
but recent advances have made it possible to explore 
the ensemble of paths \cite{MBFO11}.
For instance, an efficient path sampling has been 
proposed previously \cite{FSK10}
with the use of temperature replica exchange \cite{HN96,Hansmann97,SO99}.
The computational advantage of the action-based methods is that the
time step for the action integral can be taken to be as large as 
100 fs (or more),
which is much larger than a typical MD step ($\simeq$ 1fs).
Moreover, it is straightforward to carry out such calculations in parallel
with respect to each time step and each path.
This will certainly match the advantage of recent computational environment
with hyper-parallel architectures such as GPGPU and supercomputer.
Because of these computational demands, the ensemble of paths 
has not been easily obtained, but the fluctuations of paths 
as well as configurations should be important for biomolecules,
which is worth pursuing to understand biological functions.

In this paper, we suggest to enhance the efficiency 
of the OM action-based path sampling using the 
multiscale enhanced sampling (MSES) devised by 
Moritsugu and coworkers \cite{MTK10}.
The MSES method has been successfully 
applied to the calculation of 
free energy profiles of a small peptide (chignolin) \cite{MTK10},
intrinsically disordered protein (sortase) \cite{MTK12a}, and 
a protein-protein complex \cite{MTK12b}.
In the original MSES, the system of interest is the 
canonical ensemble for the atomistic degrees of freedom, 
which will be called ``MM DoF''.
We assume a coupling between the MM DoF and 
a reference coarse-grained DoF, which will be 
called ``CG DoF'', mimicking the typical 
behavior of the MM DoF in an economical manner.
The total Hamiltonian of the combined system is defined as 
\begin{equation}
H=H_{\rm MM}+H_{\rm CG}+ \lambda H_{\rm MMCG}
\end{equation}
where $H_{\rm MM}, H_{\rm CG}$ 
and $\lambda H_{\rm MMCG}$ 
represent the Hamiltonians of the MM and CG DoF, 
and the interaction between the MM and CG DoF, 
respectively.

 
Assuming that the CG DoF is able to move around the CG configuration 
space quickly, the MM DoF will be dragged accordingly 
for a finite value of $\lambda$, which leads to
an efficient search of the MM configurations.
However, our goal is to obtain the canonical ensemble with respect to MM-DoF,
which is proportional to $\exp(-\beta H_{\rm MM}^{})$,
in the absence of the MM-CG interaction, i.e. $\lambda=0$.
To meet both of these ends, we use the Hamiltonian replica 
exchange \cite{HRE} among
the combined MM-CG systems with different $\lambda$ values.
In other words, we simulate a replicated set of the combined MM-CG 
systems with $\lambda_1^{} > \lambda_2^{} \cdots \lambda_{N_{\rm rep}^{}}$, 
where $N_{\rm rep}^{}$ is the number of replicas, and they are 
exchanged from time to time according to
the Metropolis criterion \cite{MDbook2}.
We then collect the MM ensemble with 
the smallest $\lambda_{N_{\rm rep}^{}}$ (ideally $\lambda_{N_{\rm rep}^{}}=0$),
which should be equal to the {\it unbiased} canonical MM ensemble.

Using the OM action-based methods, we can do the same as above:
We consider an extended system where MM and CG DoF are coupled, 
and the OM action for the coupled system is defined as 
\begin{equation}
S=S_{\rm MM}+S_{\rm CG}+\lambda S_{\rm MMCG}
\end{equation}
where 
$S_{\rm MM}, S_{\rm CG}$ are the OM actions for the MM and CG DoF,
and $\lambda S_{\rm MMCG}$ the interaction between them.
Using $\lambda$ as the exchange parameter, 
the ``canonical weight'' for a MM path 
$\propto \exp (-\beta S_{\rm MM})$ can be retrieved
with the same logic as above.
Note that in the formulation of MSES there is a subtle issue 
concerning the nonequilibrium nature of the combined system 
due to the coupling to different temperatures \cite{MTK10}, 
but this problem is circumvented by using the idea of 
``scaling'' (see Appendix C). 
As shown below, when we only use the ``conventional'' OM method 
for the MM DoF (without any coupling to the CG DoF), 
we need a longer computation time for statistical convergence.
Hence it is indispensable to use this type of method to enhance 
path sampling, especially when dealing with large molecular systems.

This paper is organized as follows.
In Sec.~\ref{sec:methods}, 
we introduce the Onsager-Machlup action principle 
and formulate the multiscale enhanced sampling 
for path space using the OM action (MSES-OM). 
We also mention the model polymer system devised by 
Micheletti and coworkers \cite{TMCM06,MBL08}, 
which is employed in the 
following calculations.
In Sec.~\ref{sec:result}, 
we investigate the numerical performance of the MSES-OM method 
using the model polymer.
We also examine the numerical accuracy of the method 
by comparing with the results using the conventional transition path sampling.
In Sec.~\ref{sec:summary}, 
we summarize the paper and discuss the 
further aspects of the MSES-OM method.

\section{Formulation}
\label{sec:methods}

\subsection{Onsager-Machlup action for multi-dimensional systems}

Let us consider the case where the dynamics of the MM system 
is described by an overdamped Langevin equation,
\begin{equation}
\dot{x}_{\alpha}=-\frac{1}{\gamma}_{\alpha}\frac{\partial U}{\partial x_{\alpha}}
+\sqrt{2D_{\alpha}} \eta_{\alpha}(t)
\label{eq:Langevin}
\end{equation}
where $x_{\alpha}$ is mass-weighted coordinates, 
$U$ a potential energy,
$\gamma_{\alpha}$ an intrinsic friction coefficient,
$D_{\alpha}$ an diffusion coefficient, and 
$\eta_{\alpha}(t)$ a Gaussian-white noise satisfying 
$\langle \eta_{\alpha}(t)\eta_{\alpha'}(t') \rangle =\delta_{\alpha \alpha'}\delta (t-t')$.
As Eq.~(\ref{eq:Langevin}) is a stochastic differential equation,
the dynamics is characterized by the probability (``weight'')
of generating a path (or trajectory) $x(t)$.
According to Onsager and Machlup \cite{OM1,OM2},
the weight is proportional to $\exp(-\beta S[x(t)])$ 
where $S[x(t)]$, which is called the Onsager-Machlup (OM) action,
is a functional of $x(t)$.
In numerical calculations, the OM action is approximated by 
its discretized form as  
\begin{equation}
S[x(t)]=
\frac{\Delta U}{2}
+\sum_{i=1}^{N_{\rm bead}-1}
\left[ 
\sum_{\alpha=1}^{M}
\frac{\omega_{\alpha}^2}{2} (x_{\alpha,i+1}-x_{\alpha,i})^2 
+V_{\rm eff}( \{ x_{\alpha,i} \} )
\right] 
\label{eq:OMHam}
\end{equation}
where $N_{\rm bead}$ is the number of discretized segments of a path,
or ``beads'',
$M$ the number of the MM DoF (usually $3 \times $ number of atoms),
$\Delta U = U(x_{\rm fin})-U(x_{\rm ini})$ 
the potential difference 
between the initial and final states, 
and the effective ``potential'' is defined as 
\begin{equation}
V_{\rm eff}( \{ x_{\alpha} \})
\equiv 
\sum_{\alpha=1}^{M}
\left[ \frac{1}{8 \omega_{\alpha}^2}
U_{x_{\alpha}}^2
-\frac{k_B T}{4 \omega_{\alpha}^2}
 U_{x_{\alpha} x_{\alpha}}
\right]
\label{eq:veff}
\end{equation}
with 
\begin{equation}
\omega_{\alpha} = \sqrt{\frac{\gamma_{\alpha}}{2 \Delta t_{\rm OM}}}.
\label{eq:omega}
\end{equation}
This is an effective frequency determined by the friction and 
the time interval for the path discretization $\Delta t_{\rm OM}$.
Here we used the notations  
$U_x$ and $U_{xx}$ to represent the first and second derivatives of 
$U$ with respect to $x$.


To sample path space using the above OM action, 
we have employed the position-Verlet \cite{MDbook1,MDbook2} 
type scheme to integrate the Langevin equations, i.e.,
\begin{eqnarray}
\tilde{x}_{\alpha,i} 
&=&
x_{\alpha,i} + \frac{p_{\alpha,i}}{\tilde{m}_i} \frac{\Delta \tau}{2},
\label{eq:EOM1}
\\
p_{\alpha,i}' &=& 
(1-\tilde{\gamma} \Delta \tau/2)
p_{\alpha,i} - \left. 
\frac{\partial S}{\partial x_{\alpha,i}} \right|_{x_{\alpha,i}=\tilde{x}_{\alpha,i}} \Delta \tau  
+\sqrt{ 2 \tilde{\gamma} k_B T \tilde{m}_i \Delta \tau} R_{\alpha,i}(t),
\label{eq:EOM2}
\\
x_{\alpha,i}' &=& 
\tilde{x}_{\alpha,i} + \frac{p_{\alpha,i}'}{\tilde{m}_i} \frac{\Delta \tau}{2},
\label{eq:EOM3}
\end{eqnarray}
where $p_{\alpha,i}$, $\tilde{m}_i$, $\tilde{\gamma}$ are 
the fictitious momenta, mass, and friction coefficient,
which are similar to those for the imaginary-time 
path integral simulations \cite{Tuckermanbook}.
$R_i(t)$ is the Gaussian random variable with zero average and 
unit variance.
Note that the time step to integrate the above equations of motion 
$\Delta \tau$ has nothing to do with the time step for the 
discretization of the OM action $\Delta t_{\rm OM}$.
The numerical integration of these equations result in
the canonical distribution $\exp(-\beta S)$.
However, if the path space is huge, 
which is usually the case for complex systems,
an efficient algorithm to sample the path space 
is required to reduce the computational load.
This is the reason why we will introduce 
the MSES method \cite{MTK10,MTK12a,MTK12b}.

\subsection{Multiscale enhanced sampling for path space}
\label{sec:OM-REM}

In the formulation of MSES \cite{MTK10}, 
Moritsugu {\it et al.} introduced 
several replicas with different coupling parameters 
$\lambda_1, \lambda_2, \cdots$, 
and exchange $m$-th and $n$-th parameters during the simulation 
of the combined system according to the probability
\begin{equation}
p_{mn}= \min (1, \exp \Delta_{mn})
\end{equation}
where
\begin{equation}
\Delta_{mn}
=\beta (\lambda_m-\lambda_n)(H_{{\rm MMCG},m}-H_{{\rm MMCG},n})
\end{equation}
and $\beta$ is the inverse temperature of the MM DoF.
This is an application of the Hamiltonian 
replica exchange method \cite{HRE} to the combined MM-CG system.
The MM canonical ensemble is obtained from that of the 
combined MM-CG system with the smallest $\lambda$.

\begin{figure}
\hfill
\begin{center}
\includegraphics[scale=0.8]{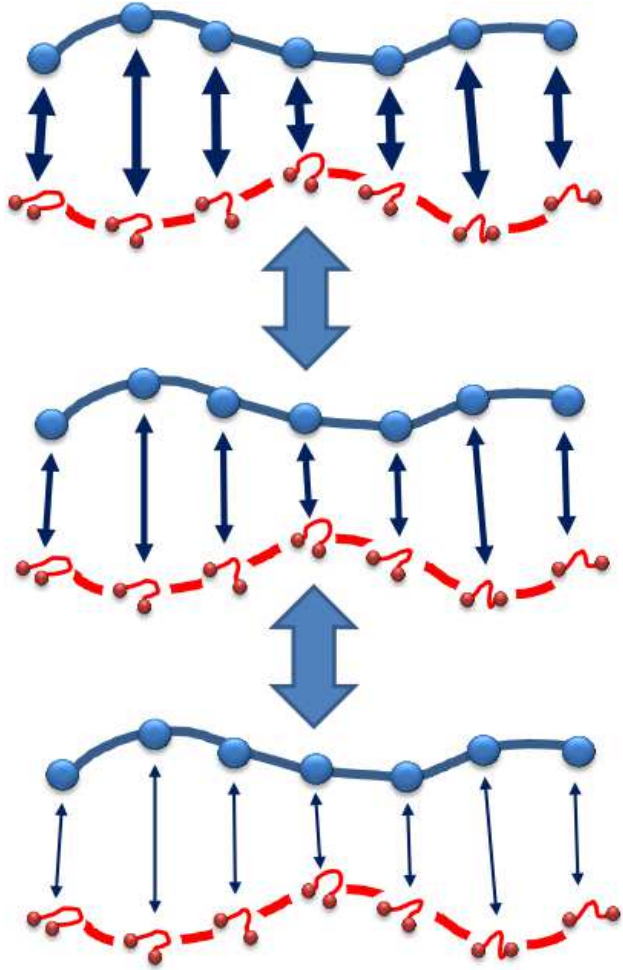}
\end{center}
\caption{\baselineskip5mm
Schematic picture of the proposed multiscale enhanced 
sampling for path space using the OM action (MSES-OM).
The blue (red) beads represent CG (MM) DoF, and the 
arrows between them with different thickness represent 
different interaction strengths. The 
thick arrows between the combined MM-CG systems 
schematically represent 
``exchange'' due to Hamiltonian replica exchange
(actually the coupling parameters are exchanged).
}
\label{fig:schematic}
\end{figure}

This formalism can be immediately applied to the 
problem of path sampling using the OM action.
In this case the exchange 
probability is determined by 
\begin{equation}
\Delta_{mn}
=\beta (\lambda_m-\lambda_n)(S_{{\rm MMCG},m}-S_{{\rm MMCG},n})
\end{equation}
where the interaction OM action can be chosen as 
\begin{equation}
\label{eq:int}
S_{\rm MMCG}= 
\frac{1}{2} \sum_{i=1}^{N^{\rm CG}_{\rm bead}} (\theta_i(\{ x_{\alpha,i} \}) -x_i^{\rm CG})^2
\end{equation}
where $\theta_i$ represents collective variables 
calculated from MM DoF $\{ x_{\alpha,i} \}$, 
$x_i^{\rm CG}$ is the corresponding 
CG variables, and $N^{\rm CG}_{\rm bead}$ is the 
number of beads to discretize the CG path. 
The schematic picture of our MSES-OM method 
is shown in Fig.~\ref{fig:schematic}.
The basic strategy here is that 
the CG variables (blue, upper) move rather freely in path space 
to ``guide'' the MM variables (red, lower).

Here it is important to notice on the scaling of the 
number of replicas required as the system size increases.
As emphasized by Moritsugu {\it et al.} \cite{MTK10}, 
the interaction ``Hamiltonian''
scales with the number of CG DoF not with that of MM DoF
because it contains only CG DoF and the projected 
MM DoF onto the CG space.  
Hence this method does not suffer from the problems 
caused by the number of MM DoF as in the case of temperature 
replica exchange \cite{Hansmann97,SO99}.
Furthermore, in this study, the number of 
CG beads is smaller than that of MM beads,
i.e., 
$N^{\rm CG}_{\rm bead}$ (=240) is ten times smaller than $N_{\rm bead}$
(=2400), reducing the cost of calculations.



In order for MSES to work efficiently, 
the CG DoF must move quickly and dominate the dynamics 
of the combined system. This situation is achieved, for example, when 
the total mass of the CG system is comparable to or larger than 
the MM system {\it and} the temperature of the CG system is
much higher than that of the MM system. 
We thus  
take the CG mass $=20 m_0$ and 
CG temperature $=80 k_B T$, whereas $m_0$ is the ``bead'' mass of 
the model polymer (see Appendix A) 
and the temperature for the MM DoF is $k_B T \simeq 300$K.
 (Note that this CG mass is 
not a real mass because this is just introduced 
to carry out the ``artificial'' OM dynamics for the CG DoF.)
Because two different temperatures 
are set for the MM and CG DoF, 
the resulting ensemble seems to exhibit a strongly 
nonequilibrium behavior. 
However, the desired information 
can be retrieved from the ensemble with the 
smallest coupling parameter $\lambda_{N_{\rm rep}}$. 
The subtle point to use different temperatures 
for the MM and CG DoF are summarized in Appendix C.

\subsection{Model polymer}
\label{sec:model}

In this section, we briefly explain the 
model system we use for path sampling, 
which is a coarse-grained polymer studied by
Micheletti and coworkers \cite{TMCM06}.
The potential energy of this model, 
$U_1+U_2+U_3$, 
consists of the Asakura-Oosawa (AO) type 
interaction $U_3$ and 
the finitely extensible elastic chain $U_1+U_2$,
\begin{eqnarray}
U_1 
&=&
\epsilon_1 \sum_{i<j} e^{-a(r_{ij}-r_{ij}^0)},
\label{eq:pot1}
\\
U_2
&=&
-\epsilon_2 \sum_i \ln 
\left[ 
1- \left( \frac{r_{i,i+1}}{1.5 r_{i,i+1}^0} \right)^2
\right],
\label{eq:pot2}
\\
U_3
&=&
-\frac{\phi k_B T}{16 r^3}
\sum_{i<j}
\left(
2 \tilde{r}_{ij} 
+3 r_{ij} 
-\frac{3 \delta_{ij}^2}{r_{ij}}
\right)
\tilde{r}_{ij}^2
\Theta(\tilde{r}_{ij}),
\label{eq:pot3}
\end{eqnarray}
where $\tilde{r}_{ij}= 2r+r_{ij}^0-r_{ij}$, 
$\delta_{ij}=|R_i-R_j|$,
$r_{ij}^0=R_i+R_j$, $\Theta(x)$ is the Heaviside function
(=1 for $x \geq 0$ and 0 for $x<0$), 
and the friction coefficient is 
given by
\begin{equation}
\gamma_i= 6 \pi \eta R_i (1+2.5 \phi).
\end{equation}
The parameter values we employ here 
are almost the same as in the original 
paper \cite{TMCM06} except that
we made $a$ parameter four times larger,
that is, we used a softer core.

Micheletti {\it et al.}
used the AO interaction to model the crowding effect 
due to RNA and proteins in a cell, 
and studied the (un)looping kinetics of the model polymer 
by changing the size of the system. They found 
that the mean-first passage time (MFPT) for the unlooping dynamics 
does not depend on the 
size of the system, whereas the MFPT for the looping dynamics
does. In this paper, we employ the 
smallest number of beads (five), and 
study the unlooping dynamics, which takes place 
with $\sim$ 3 $\mu$s time scale \cite{TMCM06}.

For path sampling using the OM action,
we need to evaluate the effective potential 
defined in Eq.~(\ref{eq:veff}) and its derivatives
because we use the equations of motion, Eqs.~(\ref{eq:EOM1})-(\ref{eq:EOM3}).
Because there are only two-body interactions in $U_1+U_2+U_3$, 
this is easily evaluated analytically.
On the other hand, for path sampling using the conventional transition 
path sampling (TPS), we integrate the overdamped 
Langevin equation (\ref{eq:Langevin}) for this model
from many different initial paths.

Although this is a coarse-grained model, 
in this paper, we call this model as an ``atomistic'' model
or MM DoF because our purpose here is to illustrate our method.
The CG DoF coupled to this MM DoF will be introduced just below.

\subsection{CG model for the model polymer and the CG variable}
\label{sec:CG}

For the MSES-OM method,
we need to introduce CG variables and its OM action.
For the model polymer introduced above, 
we take the conventional end-to-end distance $z$ 
as a CG variable, and the ``potential'' function 
is introduced as  
\begin{eqnarray}
U^{\rm CG} &=& \alpha \left[ 
\frac{U_1^{\rm CG}+U_2^{\rm CG}}{2}
-\sqrt{\left(\frac{U_1^{\rm CG}-U_2^{\rm CG}}{2}\right)^2+\epsilon^2}
\right],
\label{eq:CGpot1}
\\
U_1^{\rm CG} &=& \frac{1}{2} k_1 (z-z_1)^2 + v_1,
\label{eq:CGpot2}
\\
U_2^{\rm CG} &=& \frac{1}{2} k_2 (z-z_2)^2 + v_2,
\label{eq:CGpot3}
\end{eqnarray}
which is a conventional empirical valence-bond type 
potential \cite{MTK10}. 
In the previous study, it is shown that 
this ``single'' order parameter can nicely 
describe the unlooping process of the model 
polymer \cite{TMCM06}, so we also use this 
single variable as the CG DoF.
The values of the CG parameters are summarized in Appendix B.



\section{Results}
\label{sec:result}

\subsection{Numerical stability of OM dynamics}



It is known that the model polymer described in Sec.~\ref{sec:model}
has two free energy minima with respect to the end-to-end distance
(this is a basic observation when we build a CG model in Sec.~\ref{sec:CG}),
one at the looped configuration, $z=24$ nm, and the other at the
unlooped configuration, $z=45$ nm. 
There is a barrier between them, which are 
characterized by the barrier height with
$\simeq$ 0.4 kcal/mol (looped to barrier top) and 
$\simeq$ 0.9 kcal/mol (barrier top to unlooped).
In order to study the unlooping process,
the starting point of the OM path is fixed at $z=24$ nm
while the end point is made free to move 
around $z=45$ nm (see Fig.~\ref{fig:polymer}).
From an independent molecular dynamics run,
it has been confirmed that we need the trajectory of about 30 $\mu$s
to get the information on the barrier crossing.
Therefore, the length of OM path has been taken to be
$N_{\rm bead}^{}\Delta t_{\rm OM}^{} \simeq 40\ \mu$s
where the OM step size and the number of beads are
taken as $\Delta t_{\rm OM}^{} = 10\ t_0^{} \simeq 16$ ns
(with $t_0^{}$ being the unit of time scale, see Appendix A)
and $N_{\rm bead}^{} = 2400$, respectively.
The value of $\Delta t_{\rm OM}^{}$ may be taken as small as
possible, but due to our experience $\Delta t_{\rm OM}^{}$
is reasonably small as long as the fluctuation of the action
is balanced between the bead spring and potential contributions,
i.e. each in the second term in the right hand side of Eq.~(\ref{eq:OMHam}).
We find the sampling procedure unstable when 
taking $\Delta t_{\rm OM}^{}$ too large,
say $\Delta t_{\rm OM}^{}= 1000\ t_0^{}$, when the fluctuation 
of the potential contribution is much larger than that of 
the bead spring contribution.

\subsection{Interaction between MM and CG DoF}

\begin{figure}
\hfill
\begin{center}
\includegraphics[scale=1.2]{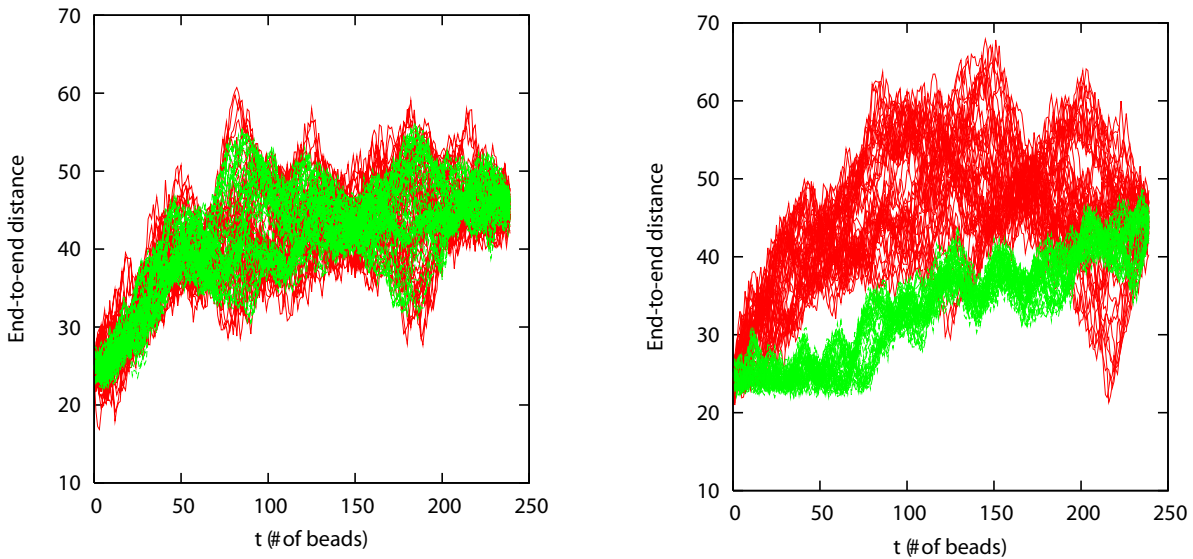}
\end{center}
\caption{\baselineskip5mm
The end-to-end distance as a function of time (or number of beads)
calculated from the OM dynamics for MM (green) and CG (red) DoF with 
the strong (left) and weak (right) coupling cases.
We overlay 10 ``snapshots'' from each OM dynamics.
There are 10 times difference in the MM and CG beads numbers,
so every one of 10 beads is shown for the MM DoF.
}
\label{fig:actions}
\end{figure}

Next we show the effect of the coupling between CG and MM paths.
In Fig.~\ref{fig:actions}, 
several CG (red) and MM (green) paths during the OM dynamics 
are drawn with two different 
coupling parameters: $\lambda = 10^{-2}$ (left) and  
$\lambda = 10^{-5}$ (right).
We can see that the MM path follows the CG path
in the strong coupling case (left), while the MM path
is more independent from the CG path in the weak
coupling case (right).

\begin{figure}
\hfill
\begin{center}
\includegraphics[scale=2.0]{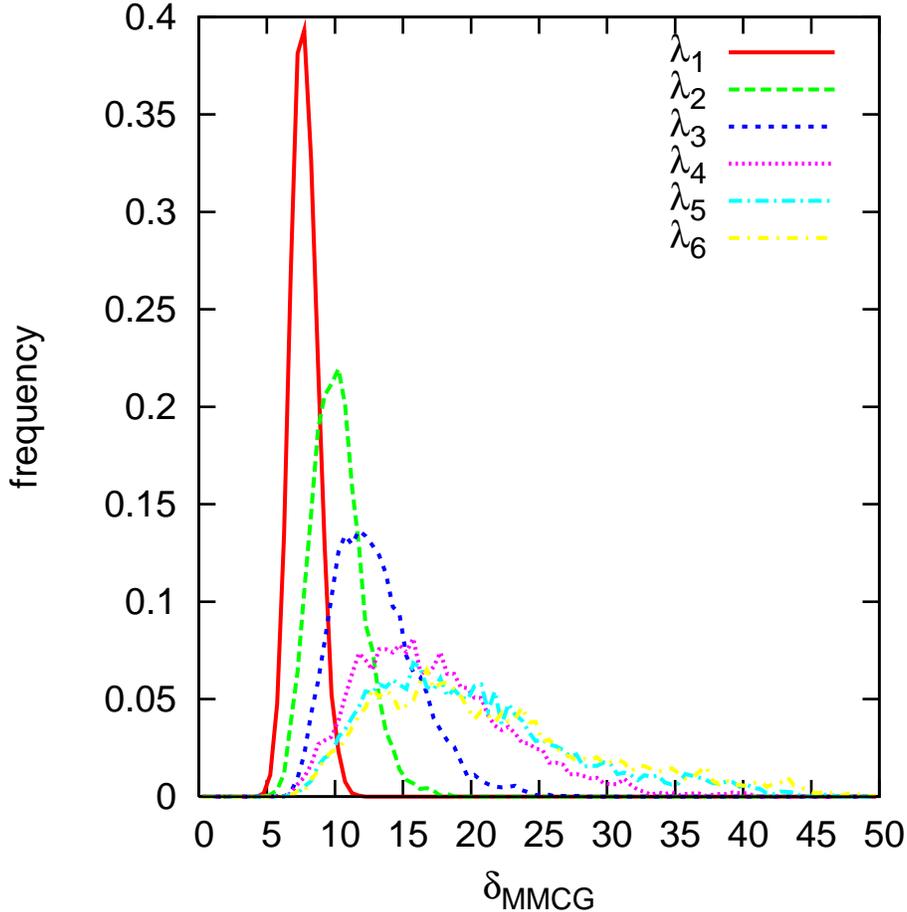}
\end{center}
\caption{\baselineskip5mm
The distribution function of the root mean square displacement 
between the MM and CG 
DoF as defined in Eq.~(\ref{eq:diff}) for different 
coupling strengths.
Note that we never used Hamiltonian replica exchange here.
}
\label{fig:dis}
\end{figure}

For a closer analysis on the correlation between
MM and CG paths,
we define the root mean square displacement (RMSD) 
between the MM and CG DoF as
\begin{equation}
\delta_{\rm MMCG}=
\sqrt{\frac{1}{N^{\rm CG}_{\rm bead}}
\sum_{i=1}^{N^{\rm CG}_{\rm bead}} (\theta_i(\{ x_{\alpha,i} \})-x^{\rm CG}_i)^2}
=\sqrt{\frac{2 S_{\rm MMCG}}{N^{\rm CG}_{\rm bead}}}.
\label{eq:diff}
\end{equation}
The histograms for this quantity over 10$^6$ OM steps 
are shown in Fig.~\ref{fig:dis} for different coupling 
strengths ($\lambda$'s).
We take 6 different values of $\lambda_i$ between 
10$^{-2}$ and 10$^{-5}$: $\lambda_i = 0.01 \times 0.25^{i-1}$ 
($i=1,\cdots,6$) for 6 different replicas.
For the largest $\lambda=\lambda_1$ (=10$^{-2}$),  
$\delta_{\rm MMCG}$ fluctuates around $\sim$ 7, 
but for the smallest $\lambda=\lambda_6$ (=10$^{-5}$),  
the distribution spreads from 5 to 50, 
indicating that the correlation is small between the MM and CG DoF.
From this figure, we assume that the paths with $\lambda_6$
is close to those with null coupling ($\lambda=0$), 
because the two distributions 
with $\lambda_5$ and $\lambda_6$ are almost the same.


\subsection{First passage time distribution and comparison to TPS}

\begin{figure}
\hfill
\begin{center}
\includegraphics[scale=0.7]{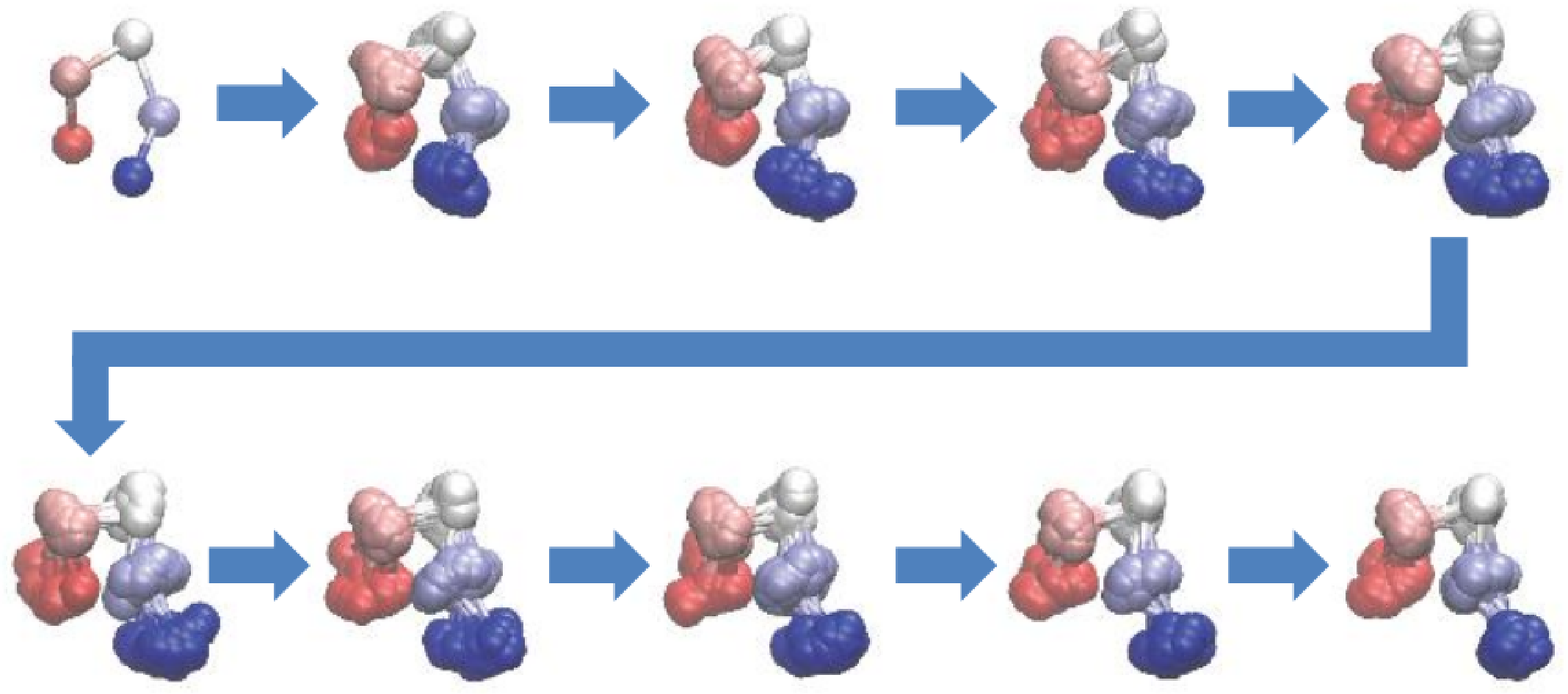}
\end{center}
\caption{\baselineskip5mm
Unbiased path ensemble calculated by the MSES-OM method.
A hundred of OM trajectories are superimposed, 
indicating the fluctuation of the path ensemble.
The ``time difference'' between the neighboring snapshots 
is about 4 $\mu$s, corresponding to 240 beads.
}
\label{fig:polymer}
\end{figure}

We carried out a MSES-OM calculation 
with 10$^6$ OM steps.
The path data were saved every 100 OM steps, 
and 10$^4$ data sets were collected.
In the MSES-OM calculation, 
the replica exchange was attempted every 10 OM steps,
and the acceptance ratio was found to be about 30 \%.
In Fig.~\ref{fig:polymer}, we show the overlay of snapshots 
of 100 OM paths from the calculation.
The time course is indicated by arrows, 
and 
the ``time difference'' between the neighboring snapshots 
is about 4 $\mu$s, corresponding to 240 beads.
We can see that the distribution of paths grows 
during the time course, but it is hard to 
tell what is happening here.

\begin{figure}
\hfill
\begin{center}
\includegraphics[scale=1.0]{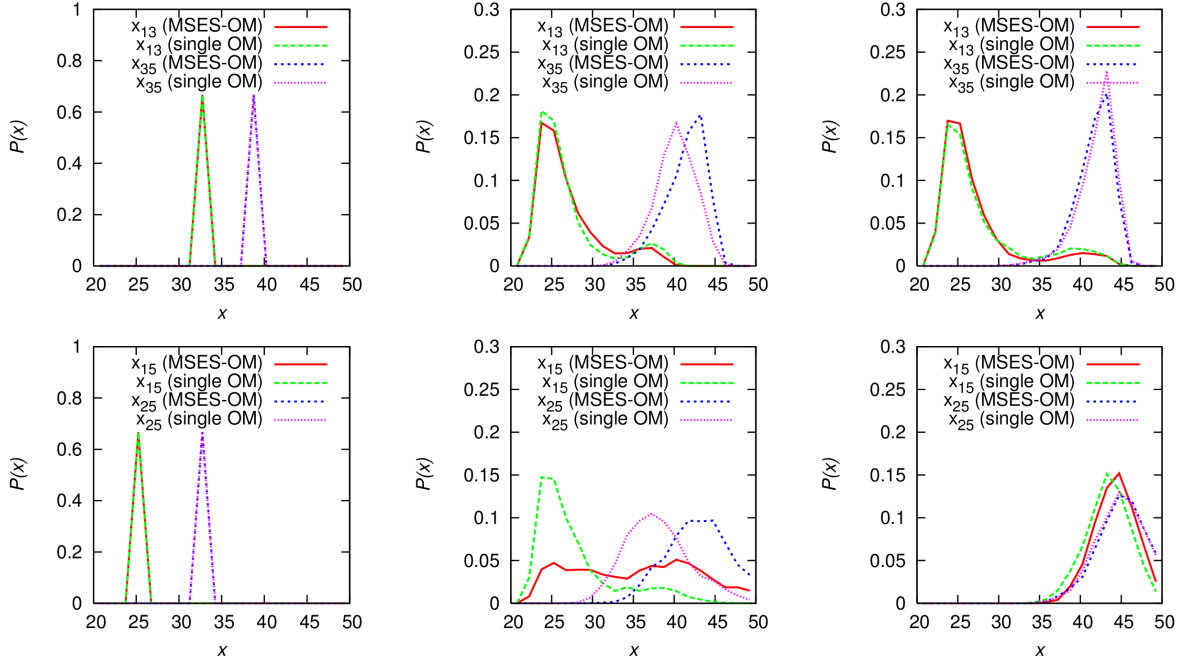}
\end{center}
\caption{\baselineskip5mm
Distribution functions for different distances,
$x_{13}, x_{35}$ for the upper panel and 
$x_{15}, x_{25}$ for the lower panel,
of the model polymer calculated by the single OM and the MSES-OM method.
The left, middle, and right panels correspond
to the distributions at 
$t=0$, $1233 \Delta t_{\rm OM}$, and $2323 \Delta t_{\rm OM}$,
respectively.
}
\label{fig:polymer2}
\end{figure}

In Fig.~\ref{fig:polymer2}, 
we show the time evolution of the distribution with respect
to the distances between ``beads'' $i$ and $j$ denoted by $x_{ij}^{}$.
For comparison, we also showed the result using the conventional OM calculation 
without Hamiltonian replica exchange, which is 
called ``single OM'' in the figure.
%
%
We can see that during the transition 
the end-to-end distance, $x_{15}^{}$,
is broadly distributed among the two basins (middle-bottom panel).
In contrast, the distances distributions 
for $x_{13}^{}$ and $x_{35}^{}$
are already close to those at the final time,
%
indicating that these variables, $x_{13}^{}$ and $x_{35}^{}$, 
first equilibrate before the end-to-end distance does.
The difference between the single OM and MSES-OM methods is found
in the distributions for $x_{15}^{}$ and $x_{25}^{}$ during 
the transition.
This clearly shows the efficiency of our MSES-OM method over
the single OM method for path sampling.

\begin{figure}
\hfill
\begin{center}
\includegraphics[scale=2.0]{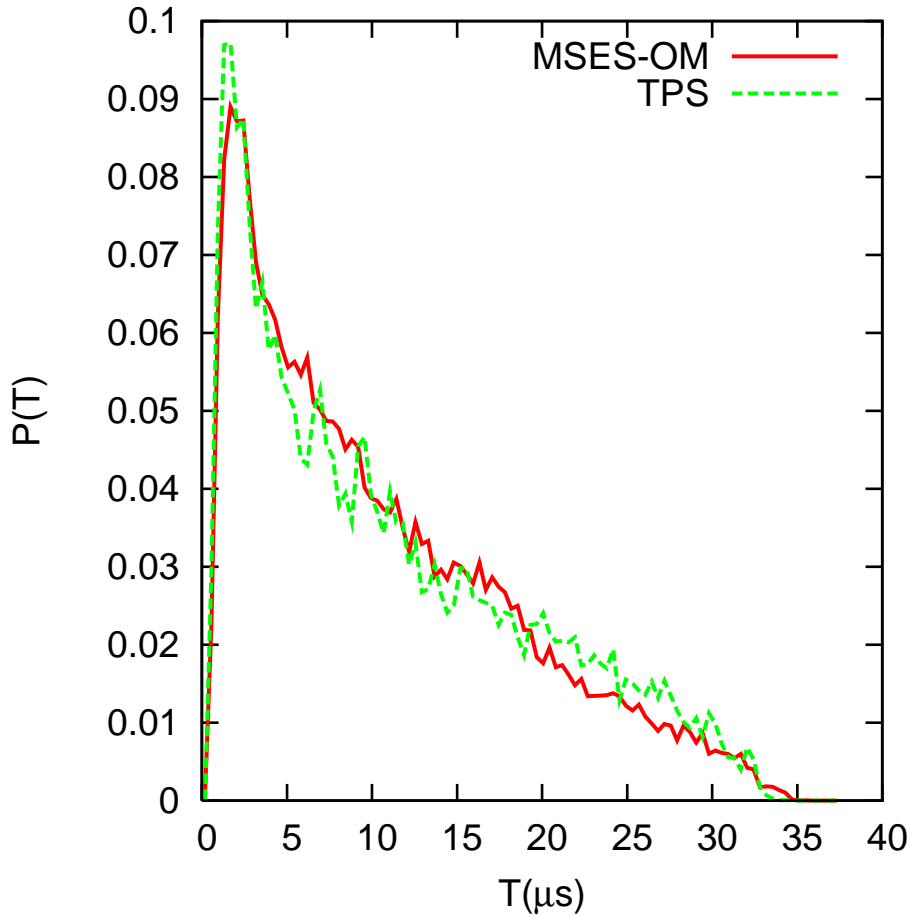}
\end{center}
\caption{\baselineskip5mm
First passage time distributions (FPTDs) for the model 
polymer calculated by TPS (green) and by the MSES-OM method (red).
}
\label{fig:FPTD}
\end{figure}

Finally we show the first passage time distribution (FPTD) 
in Fig.~\ref{fig:FPTD} \cite{HCB10,ZJZ07}.
This is defined as a histogram of the elapsed times 
in moving from one place to the other. 
In this work we monitor the end-to-end 
distance $z=x_{15}$ and measure the time interval 
where the system evolves from $z_1^{} = 28$ to $z_2^{} = 35$,
which correspond to two locations before and after the 
barrier crossing (not free energy minima) \cite{ZJZ07}.
We can see that the barrier crossing occurs most likely in
2 $\mu$s, and completes within 35 $\mu$s.
In order to have a converged result for Fig.~\ref{fig:FPTD}, 
we repeated the MSES-OM calculation
10 times from different initial conditions, so the 
number of path in the ensemble is 10$^5$.

As a reference, we have also carried out TPS calculations 
using the Crooks-Chandler algorithm \cite{CC01} for the same system.
As can be seen, the FPTD is in reasonable 
agreement with the MSES-OM result.
To obtain Fig.~\ref{fig:FPTD}, 
we used 100 independent TPS runs with 10$^4$ TPS steps each.
The data is saved every 10 TPS steps, and 10$^5$ data sets 
are collected in total.
Thus, the amount of data for TPS calculation and the MSES-OM 
calculation is comparable.
The statistical accuracy in Fig.~\ref{fig:FPTD} seems to be similar
between the MSES-OM and TPS methods, at least for the present system.

However, the comparison of computational effort is difficult in general,
since the OM methods requires the higher-order derivatives of the potential
whereas the parallelized calculation can be done efficiently.
We can at least say that OM method would be of use in studying slow and diffusive
processes such as conformational changes of large molecules and protein folding.
There should be much room for further improvement of the OM method such as
the dominant pathway \cite{Orland09}, the discretization in length \cite{EGC02},
and the use of the efficient hessian calculation \cite{EH08}.

\section{Concluding remarks}
\label{sec:summary}


To efficiently sample huge path space of complex systems, 
in this paper we proposed a multiscale path sampling method 
based on the Onsager-Machlup (OM) action combined with 
multiscale enhanced sampling (MSES).
We applied this MSES-OM method to a model polymer with 
the Asakura-Oosawa interaction, and showed the 
effectiveness of our method, which is comparable 
to conventional transition path sampling (TPS) with 
many initial paths.


We have shown that the MSES method can be easily combined 
with the OM action.
This is because the ensemble of stochastic paths
subject to overdamped Langevin equation is isomorphic to the canonical ensemble
of a polymer of beads connected by harmonic springs.
Note that the path obtained by the OM or MSES-OM method 
contains the nonequilibrium information 
such as the first passage time for barrier crossing.
This is conceptually different from the thermodynamic 
analysis of barrier crossing, i.e., transition state 
theory based on the free energy landscape.
In order to apply this method to complex molecular systems, which is obviously 
the next step, it is indispensable to make a great use of 
parallel calculation techniques.
For a future work, the authors are interested in studying the 
conformational change of adenylate kinase \cite{MFTFMK12}, 
for which the minimum free energy path was recently 
elucidated by the on-the-fly string method \cite{ME07}.

Another direction to pursue is the Baysian-type 
inference of the model parameters using the OM action \cite{MBL08,Harada}.
It is known that the formulation of the 
dynamic data assimilation \cite{data_assimilation} 
is almost the same as our MSES-OM method, 
where the CG DoF 
corresponds 
the direct variables to be observed, and 
the MM DoF to the hidden variables of the system studied.
Hence the MSES-OM formalism should be used for 
the parameter estimation of general dynamical system problems.


Finally we mention a possible extension of this approach 
to quantum systems. 
Since the centroid variable of a quantum 
particle has been used in TPS \cite{AS09} and in the string method \cite{SF12}, 
we expect that we can use the centroid variable in the OM-action 
based methods as well.
This will be an extension of the OM path sampling 
based on the Feynman-Hibbs formula which was 
recently suggested by Faccioli and coworkers \cite{BGF11,BF12}.



\begin{acknowledgements}

One of the authors (HF) is grateful to 
Daniel M. Zuckerman, 
Luca Maragliano, Pietro Faccioli,
Yukito Iba, Shin-ichi Sasa, Koji Hukushima, 
Tohru Terada, Ikuo Fukuda, Yasuhiro Matsunaga for useful discussions.
This research was supported by Research and Development of 
the Next-Generation Integrated Simulation of Living Matter,
a part of the Development and Use of the Next-Generation 
Supercomputer Project of the Ministry of Education,
Culture, Sports, Science and Technology (MEXT).
The computations were partly performed on the RIKEN Integrated 
Cluster of Clusters (RICC).

\end{acknowledgements}

\begin{appendix}

\section{Units for dimensionless variables}

We here summarize the dimensionless units 
which were used in this paper.
We convert all the variables (mass, length, energy, time, friction 
coefficient) into dimensionless ones 
by choosing
\begin{eqnarray}
m_0 
&=&
\rho \cdot  
\frac{4}{3} \pi R^3
=\frac{4}{3} \pi \times
1.35 \, {\rm g/cm}^3 \times (12.5 \, {\rm nm})^3
=1.10 \times 10^{-20} \, {\rm kg},
\\
l_0
&=&
10^{-9} \, {\rm m} 
= 1 \, {\rm nm},
\\
\epsilon_0 
&=&
k_B T = 1.38 \times 10^{-23}\, {\rm J/K} \times 300 \, {\rm K},
\simeq 
4.14 \times 10^{-21} \, {\rm J},
\\
t_0
&=&
l_0 \sqrt{\frac{m_0}{\epsilon_0}}
=1.63 \times 10^{-9} \, {\rm s}
=1.63 \, {\rm ns},
\\
\gamma_0
&=& m_0/t_0 = 6.75 \times 10^{-12} \, {\rm kg/s}.
\end{eqnarray}
The order of the friction coefficient is given by
\begin{eqnarray}
\gamma_i &=& 
6 \pi \eta R_i = 
6 \pi \times 0.05 \times 10^{-3} \,{\rm g/cm/s} \times 12.5 \, {\rm nm}
\nonumber
\\
&\simeq& 1.18 \times 10^{-9} \, {\rm kg/s} 
= 175 \gamma_0.
\end{eqnarray}
If we take as a time step for 
the Langevin dynamics: $\Delta t = t_0/100=16.3$ ps, 
which is a reasonable choice, 
the jump magnitude due to the random force is 
\begin{equation}
\sqrt{\frac{2 k_B T}{\gamma_i} \Delta t} 
=\Delta L \simeq 0.01 \, {\rm nm}
\end{equation}
which seems to be reasonable as well for this model polymer system.



\section{Parameters for models and simulations}

\subsection{Parameters for the MM DoF}

The 
parameters used for the model polymer 
defined in Eqs.~(\ref{eq:pot1})-(\ref{eq:pot3})
are 
$a = 4 \, {\rm nm}^{-1}, 
\epsilon_1 = 0.24 \epsilon_0, 
\epsilon_2 = 70 \epsilon_0,
r = 2.5 \, {\rm nm},
\phi = 0.15, 
R_i = 12.5 \, {\rm nm},
\eta = 5 \, {\rm cP},
\rho = 1.35 \, {\rm g/cm}^3$,
which are the same as in \cite{TMCM06} except the value of 
$a$. 

\subsection{Parameters for the CG DoF}

The parameters used for the CG model
defined in Eqs.~(\ref{eq:CGpot1})-(\ref{eq:CGpot3})
are 
$k_1=0.8, k_2=0.02, z_1=24, z_2=46, v_1=3.5, v_2=-2.0, 
\epsilon=5.0,\alpha=0.7$.
The parameters for the OM action are 
$\Delta t_{\rm OM}^{\rm CG}=100, \zeta^{\rm CG}=333, 
D^{\rm CG}=0.003, N^{\rm CG}_{\rm bead}=240$
such that the total duration becomes the same as the MM case and 
$k_B T=1.0=D \zeta$. (However, this matching of the timescales 
for both DoF might not be necessary.)
The value of the diffusion constant $D=0.003$ was a little bit 
smaller than that of Fig. 4 in \cite{MBL08}. 
In any case, we can estimate this value from a short-time simulation.

\section{On the use of different temperatures for MM and CG DoFs}

At first sight, if we use different temperatures coupled to different 
parts of the system, the system is in a nonequilibrium state 
which cannot be described by the canonical distribution.
This is the case in our path ensemble as well as the configuration
ensemble in the original paper of MSES \cite{MTK10}.
However, this problem can be bypassed in the following way.

Because the formalism is the same for both configuration and path 
sampling, we employ the 
original setting for configuration sampling.
For simplicity we consider only a two-dimensional case with 
CG variables $(Q,P)$ and MM variables $(q,p)$, 
but this can be easily generalized.

We set the total (combined) Hamiltonian as 
\begin{eqnarray}
H &=& H_{\rm MM}(q,p)+H_{\rm CG}(Q,P)+V_{\rm MMCG}(q,Q)
\nonumber
\\
&=&
\frac{p^2}{2 m} +V_{\rm MM}(q)+ \frac{P^2}{2M} 
+ V_{\rm CG}(Q)+V_{\rm MMCG}(q,Q)
\end{eqnarray}
where $m,M$ are masses for MM and CG DoF.
To realize the canonical distribution for the 
total system, we consider the underdamped Langevin
dynamics. The equations for the original variables 
are 
\begin{eqnarray}
\frac{dp}{dt}
&=& -\gamma_{\rm MM} p -\frac{\partial V_{\rm MM}}{\partial q}
-\lambda \frac{\partial V_{\rm MMCG}}{\partial q}
+\sqrt{2 \gamma_{\rm MM} m k_B T_{\rm MM}} \eta_{\rm MM}(t),
\label{eq:Langevin1}
\\
\frac{dP}{dt}
&=& -\gamma_{\rm CG} P -\frac{\partial V_{\rm CG}}{\partial Q}
-\lambda \frac{\partial V_{\rm MMCG}}{\partial Q}
+\sqrt{2 \gamma_{\rm CG} M k_B T_{\rm MM}}  \eta_{\rm CG}(t),
\label{eq:Langevin2}
\end{eqnarray}
where $\gamma_*$ is the friction coefficient 
and $\eta_*(t)$ is the Gaussian white noise.
Note that we used the same temperature $T_{\rm MM}$ to 
realize the canonical distribution for the total system 
$\propto e^{-H/k_B T_{\rm MM}}$.

Changing the CG variables from $(Q,P)$ to $(\tilde{Q},\tilde{P})$, 
which are defined by 
\begin{equation}
\tilde{Q} =\sqrt{\alpha}Q, \tilde{P} =P/\sqrt{\alpha}, 
\end{equation}
with $\alpha=T_{\rm MM}/T_{\rm CG}$ ($\leq 1$), 
the Langevin equation for the CG variable becomes
\begin{equation}
\frac{d\tilde{P}}{dt}
= -\gamma_{\rm CG} \tilde{P} -\frac{\partial V_{\rm CG}}{\partial \tilde{Q}}
-\lambda \frac{\partial V_{\rm MMCG}}{\partial \tilde{Q}}
+\sqrt{2 \gamma_{\rm CG} M k_B T_{\rm CG}}  \eta_{\rm CG}(t).
\label{eq:Langevin3}
\end{equation}
Note that the temperature is $T_{\rm CG}$ 
in this equation instead of $T_{\rm MM}$.

To have the same $\alpha$ scaling for the 
terms proportional to $\lambda$ 
(the third terms 
in Eqs.~(\ref{eq:Langevin1}), (\ref{eq:Langevin2}), (\ref{eq:Langevin3}) 
on the right hand side),
the interaction Hamiltonian has to be
\begin{equation}
V_{\rm MMCG}(q,Q)
=\frac{\lambda}{2} (\theta(q)-\sqrt{\alpha}{Q})^2
\end{equation}
where $\theta(q)$ represents 
the transformation from the MM variable to a collective one.

Hence the above total Hamiltonian can be recast into 
\begin{equation}
H= H_{\rm MM}(q,p)+\frac{\tilde{P}^2}{2 \tilde{M}} 
+V_{\rm CG}(\tilde{Q}/\sqrt{\alpha})
+V_{\rm MMCG}(q,\tilde{Q}/\sqrt{\alpha})
\end{equation}
with $\tilde{M}=M/\alpha$. 
As a result, by simulating Eqs.~(\ref{eq:Langevin1}) 
and (\ref{eq:Langevin3}) with different temperatures, $T_{\rm MM}$ 
and $T_{\rm CG}$, we can achieve the canonical distribution for 
the total system with a single temperature $T_{\rm MM}$.
In actual calculations, we use only $\tilde{Q}$ space 
to construct the CG model, $\tilde{V}_{\rm CG}(\tilde{Q}) \equiv
V_{\rm CG}(\tilde{Q}/\sqrt{\alpha})$, 
and the interaction is 
$\tilde{V}_{\rm MMCG}(q,\tilde{Q})
\equiv V_{\rm MMCG}(q,\tilde{Q}/\sqrt{\alpha})
=\frac{\lambda}{2} (\theta(q)-\tilde{Q})^2$,
it is reasonable to take 
$\tilde{V}_{\rm CG}(\tilde{Q})$
which can mimic the original free energy landscape
calculated from $V_{\rm MM}(q)$.

\end{appendix}


\begin{thebibliography}{00}

\bibitem{MDbook1}
M.P. Allen and D.J. Tildesley,
{\it Computer Simulation of Liquids},
Oxford Science, Oxford (1987).

\bibitem{MDbook2}
D. Frenkel and B. Smit,
{\it Understanding Molecular Simulation: From Algorithms 
to Applications}, 2nd ed., Academic Press (2002).



\bibitem{Anton}
J.L. Klepeis, K. Lindorff-Larsen, R.O. Dror, and D.E. Shaw,
Curr. Opin. Struc. Biol. {\bf 19}, 120-127 (2009); 
R.O. Dror, R.M. Dirks, J.P. Grossman, H. Xu, and D.E. Shaw,
Annu. Rev. Biophys. {\bf 41}, 429 (2012).

\bibitem{amber}
http://ambermd.org/

\bibitem{gromacs}
http://www.gromacs.org/

\bibitem{charmm}
http://www.charmm.org/

\bibitem{namd}
http://www.ks.uiuc.edu/Research/namd/


\bibitem{TPS}
C. Dellago, P.G. Bolhuis, and P.L. Geissler,
Adv. Chem. Phys. {\bf 123}, 1-78 (2002). 


\bibitem{TPS2}
C. Dellago and P.G. Bolhuis, 
Top. Curr. Chem. {\bf 268}, 291-317 (2007).


\bibitem{TPS3}
C. Dellago and P.G. Bolhuis, Adv. Poly. Sci. {\bf 221}, 167 (2008).


\bibitem{FFMK09}
S. Fuchigami, H. Fujisaki, Y. Matsunaga, and A. Kidera,
Adv. Chem. Phys. {\bf 145}, 35 (2011).

\bibitem{TIS}
T.S. van Erp, D. Moroni, and P.G. Bolhuis,
J. Chem. Phys. {\bf 118}, 7762 (2003);
T.S. van Erp and P.G. Bolhuis,
J. Comp. Phys. {\bf 205}, 157 (2005).

\bibitem{PPTIS}
D. Moroni, T.S. van Erp, and P.G. Bolhuis,
J. Chem. Phys. {\bf 120}, 4055 (2004).

\bibitem{milestone}
T. Faradjian and R. Elber,
J. Chem. Phys. {\bf 120}, 10882 (2004);
A.M.A. West, R. Elber. and D. Shalloway,
J. Chem. Phys. {\bf 126}, 145104 (2007).

\bibitem{FFsampling}
R.J. Allen, D. Frenkel, and P.R. ten Wolde, 
J. Chem. Phys. {\bf 124}, 024102 (2004);
R.J. Allen, D. Frenkel, and P.R. ten Wolde, 
J. Chem. Phys. {\bf 124}, 194111 (2004).



\bibitem{Wiegelbook}
F.W. Wiegel,
{\it Introduction to Path-Integral Methods in Physics 
and Polymer Science}, World Scientific, Singapore (1986). 

\bibitem{Zuckermanbook}
D.M. Zuckerman,
{\it Statistical Physics of Biomolecules: An Introduction},
CRC Press (2010).



\bibitem{EGC02}
R. Elber, A. Ghosh, and A. Cardenas,
Acc. Chem. Res. {\bf 35}, 396 (2002). 


\bibitem{EGD01}
P. Eastman, N. Gronbech-Jensen, and S. Doniach,
J. Chem. Phys. {\bf 114}, 3823 (2001). 


\bibitem{Orland09}
P. Faccioli, M. Sega, F. Pederiva, and H. Orland,
Phys. Rev. Lett. {\bf 97} (2006) 108101;
M. Sega, P. Faccioli, F. Pederiva, G. Garberoglio, and H. Orland,
ibid. {\bf 99} (2007) 118102;
E. Autieri, P. Faccioli, M. Sega, F. Pederiva, and H. Orland,
J. Chem. Phys. {\bf 130}, 064106 (2009). 

\bibitem{MBFO11}
G. Mazzola, S.a Beccara, P. Faccioli, and H. Orland,
J. Chem. Phys. {\bf 134}, 164109 (2011). 


\bibitem{FSK10}
H. Fujisaki, M. Shiga, and A. Kidera,
J. Chem. Phys. {\bf 132}, 134101 (2010).   

\bibitem{HN96}
K. Hukushima and K. Nemoto,
J. Phys. Soc. Jpn. {\bf 65}, 1604 (1996).


\bibitem{Hansmann97}
U.H.E. Hansmann,
Chem. Phys. Lett. {\bf 281}, 140-150 (1997).

\bibitem{SO99}
Y. Sugita and Y. Okamoto,
Chem. Phys. Lett. {\bf 314}, 141-151 (1999).




\bibitem{MTK10}
K. Moritsugu, T. Terada, and A. Kidera,
J. Chem. Phys. {\bf 133}, 224105 (2010).   

\bibitem{MTK12a}
K. Moritsugu, T. Terada, and A. Kidera,
J. Am. Chem. Soc. {\bf 134}, 7094 (2012).

\bibitem{MTK12b}
K. Moritsugu, T. Terada, and A. Kidera,
unpublished.

\bibitem{HRE}
H. Fukunishi, O. Watanabe, and S. Takada,
J. Chem. Phys. {\bf 116}, 9058 (2002). 


\bibitem{TMCM06}
N.M. Toan, D. Marenduzzo, P.R. Cook, and C. Micheletti,
Phys. Rev. Lett. {\bf 97}, 178302 (2006).

\bibitem{MBL08}
C. Micheletti, G. Bussi, and A. Laio,
J. Chem. Phys. {\bf 129}, 074105 (2008).   


\bibitem{OM1}
L. Onsager and S. Machlup,
Phys. Rev. {\bf 91}, 1505 (1953).

\bibitem{OM2}
S. Machlup and L. Onsager,
Phys. Rev. {\bf 91}, 1512 (1953).

\bibitem{Tuckermanbook}
M.E. Tuckerman, {\it Statistical Mechanics: Theory and Molecular Simulation},
Oxford University Press (2010).



\bibitem{HCB10}
Z. Hu, L. Cheng and B.J. Berne, 
J. Chem. Phys. {\bf 133}, 034105 (2010).

\bibitem{ZJZ07}
B.W. Zhang, D. Jasnow, and D.M. Zuckerman,
Proc. Nat. Acad. Sci. USA, {\bf 104}, 18043 
 (2007). 

\bibitem{CC01}
G.E. Crooks and D. Chandler,
Phys. Rev. E {\bf 64}, 026109 (2001).

\bibitem{EH08}
E. Vanden-Eijnden and M. Heymann,
J. Chem. Phys. {\bf 128}, 061103 (2008). 



\bibitem{MFTFMK12}
Y. Matsunaga, H. Fujisaki, T. Terada, T. Furuta, K. Moritsugu, and A. Kidera,
PLoS Comput. Biol. {\bf 8}, e1002555-1-12 (2012).   

\bibitem{ME07}
L. Maragliano and E. Vanden-Eijnden, 
Chem. Phys. Lett. {\bf 446}, 182 (2007).



\bibitem{Harada}
M. Miyazaki and T. Harada,
J. Chem. Phys. {\bf 134}, 085108 (2011).

\bibitem{data_assimilation}
A. Apte, M. Hairer, A.M. Stuart, and J. Voss,
Physica D {\bf 230}, 50 (2007).




\bibitem{AS09}
D. Antoniou and S.D. Schwartz,
J. Chem. Phys. {\bf 131}, 224111 (2009).

\bibitem{SF12}
M. Shiga and H. Fujisaki,
J. Chem. Phys. {\bf 136}, 184103 (2012).

\bibitem{BGF11}
S. a Beccara, G. Garberoglio, and P. Faccioli,
J. Chem. Phys. {\bf 135}, 034103 (2011).

\bibitem{BF12}
L. Boninsegna and P. Faccioli,
J. Chem. Phys. {\bf 136}, 214111 (2012).






%


\end{thebibliography}
\end{document}